\documentclass[]{mn2e} \usepackage{epsf}

\title[Lightcurve Classification in Massive Variability Surveys II:
      Transients towards the LMC.]
      {Lightcurve Classification in Massive Variability Surveys II:
      Transients towards the Large Magellanic Cloud.}

\author[V. Belokurov et al.]
       {Vasily Belokurov$^{1,2}$, N. Wyn Evans$^2$, Yann Le Du$^1$ \\
       $^1$ Theoretical Physics, Department of Physics, 1 Keble Road,
      Oxford, OX1 3NP, UK \\
       $^2$ Institute of Astronomy, Madingley Rd, Cambridge, CB3 0HA, UK}

\def\spose#1{\hbox to 0pt{#1\hss}}
\def\lta{\mathrel{\spose{\lower 3pt\hbox{$\sim$}} \raise
2.0pt\hbox{$<$}}}
\def\gta{\mathrel{\spose{\lower 3pt\hbox{$\sim$}} \raise
2.0pt\hbox{$>$}}}

\begin{document} 

\label{firstpage}

\maketitle

\begin{abstract}
Automatic classification of variability is now possible with tools
like neural networks. Here, we present two neural networks for the
identification of microlensing events -- the first discriminates
against variable stars and the second against supernovae. The inputs
to the networks include parameters describing the shape and the size
of the lightcurve, together with colour of the event. The network
computes the posterior probability of microlensing, together with an
estimate of the likely error. An algorithm is devised for direct
calculation of the microlensing rate from the output of the neural
networks.  We present a new analysis of the microlensing candidates
towards the Large Magellanic Cloud (LMC). The neural networks confirm
the microlensing nature of only 7 of the possible 17 events identified
by the MACHO experiment. This suggests that earlier estimates of the
microlensing optical depth towards the LMC may have been
overestimated. A smaller number of events is consistent with the
assumption that all the microlensing events are caused by the known
stellar populations in the outer Galaxy/LMC.
\end{abstract}

\begin{keywords}
gravitational lensing -- variable stars -- data processing
\end{keywords}

\section{Introduction}

Microlensing is rare and out-numbered by stellar variability by at
least a factor of ten thousand. Despite this, the selection of
microlensing candidates in variability surveys seems straightforward at
an optimistic first glance.  Unlike almost all forms of stellar
variability, microlensing is achromatic, time-symmetric and does not
repeat. The theoretical form of the microlensing lightcurve is
well-known (e.g., Paczy\'nski 1986) and so events can seemingly be
selected by their goodness-of-fit in two passbands.

In practice, the selection of candidates is fraught with difficulties.
The lightcurves are usually sparsely sampled and noisy -- for example,
the median seeing at the site of one of the most prominent
microlensing experiment (MACHO) is $\sim 2.0''$.  More awkwardly
still, the clear-cut set of characteristics of microlensing only holds
good in the simplest case of an isolated point-mass lensing a
point-source. In fact, microlensing lightcurves may show colour
variations because of blending (e.g., Di Stefano \& Esin 1995). They
may show substantial deviations from time-symmetry because of parallax
or xallarap effects (Dominik 1998; Mao et al. 2002) or because the
lens is a binary star (e.g., Mao \& Paczy\'nski 1991; An et al. 2004).

As a consequence, the results of the microlensing experiments towards
the Magellanic Clouds by the MACHO and EROS collaborations remain
controversial (e.g., Evans 2002). From 5.7 years of data, the MACHO
collaboration identified between 13 and 17 candidates towards the
Large Magellanic Cloud (LMC) and reckoned that the optical depth is
$1.2^{+0.4}_{-0.3} \times 10^{-7}$ (Alcock et al. 2000). The first
set of 13 events comprises the most convincing candidates, whilst the
second set of 17 candidates includes an additional 4 events less
firmly established. This is in astonishing contrast to the results
reported by the EROS collaboration, who found just 3 events towards
the LMC (Lasserre et al. 2000). The two experiments are not directly
comparable as EROS monitor a wider solid angle of less crowded fields
than do MACHO. Even though EROS do not analyze their data in terms of
optical depth, it is clear that the results point to a lower value
than that claimed by MACHO. Tellingly, a similar discord prevails in
the results towards the Galactic Centre; MACHO (Alcock et al. 1997)
recorded that the microlensing optical depth to the red clump stars as
$3.9^{+1.8}_{-1.2} \times 10^{-6}$, while EROS (Afonso et al. 2003b)
found a value of $0.94 \pm 0.26 \times 10^{-6}$ at almost the same
location.  These discrepancies strongly suggest that the systematic
effects in the experiments are not yet fully understood, with
candidate selection fingered as the most likely culprit.

All this motivated Belokurov, Evans \& Le Du (2003) to introduce
neural networks as an automatic way of classifying lightcurve shapes
in massive variability surveys.  They constructed a working neural
network for identification of microlensing events and applied it to
microlensing data towards the Galactic Centre. In this paper, the
ideas and methods of analysis are extended to the variability datasets
taken towards the LMC. This is a harder problem, as the source stars
are fainter and hence the microlensing events less clear-cut. A
particular difficulty already identified by Alcock et al. (2000) is
the contamination of samples of microlensing events by supernovae in
distant galaxies behind the LMC.

\begin{table*}
\begin{center}
\begin{tabular}{l|l|c}\hline
Variable Type & Specific Examples & Number\\ \hline
Eruptive      & Pre-Main Sequence, R Corona Borealis stars & 34 \\
Pulsating     & RV Tauris, Mira, Semi-Regular variables & 595 \\
\null         & Cepheids & 372 \\
\null         & Bumpers  & 300 \\
Cataclysmic   & Supernovae, novae, recurrent novae & 45\\
Eclipsing     & \null & 135 \\
MACHO samples & \null & 531 \\
Microlensing  & \null & 1500 \\
\hline
\end{tabular}
\end{center}
\caption{Composition of the training set. There are 1500 examples of
microlensing and 2014 examples of other classes of lightcurves.  The
sources for the data are reported in the main text. }
\label{table:tset}
\end{table*}

\section{Lightcurve classification with neural networks}

Let us briefly review the main stages of a classification routine with
neural networks (see Bishop 1995 for more details). As a first step,
the lightcurves are pre-processed with the primary goal of reducing
the amount of data to be examined. Features can be extracted
automatically, for example, with the help of spectral analysis or
principle component analysis. Alternatively, we can try to
incorporate {\it a priori} information and use only those features
that are believed to quantify characteristic properties of the
lightcurve, such as shape, periodicity or colour. These features are
then normalized to provide inputs for the neural network. An optimum
choice of inputs is the key to success.

The next stage involves choosing a particular architecture for the
neural network (such as the number of hidden units or layers) and
training the network on the set of previously classified patterns of
inputs $x_i$. The logistic activation function is used and the output
neuron takes values in the range between 0 and 1. Thus, the output $y$
models the posterior probability of the variability classes (see e.g.,
Bishop 1995 or Belokurov et al. 2003). Training is performed by
minimizing the error function, which consists of the standard
cross-entropy term and the weight decay term $\alpha\sum_{i} w^2_i$,
where the sum runs over all weights $w_i$. Adjusting a hyper-parameter
$\alpha$ enables one to control the magnitude of weights and hence to
minimise any over-fitting. This can be done automatically during
training. This differs from the procedure used in Belokurov et
al. (2003), as no validation set is required and the whole of the
available data can be used as a training set. Further reduction of the
variance in network predictions can be achieved by using a committee
of networks. A very inexpensive but efficient way of introducing the
committee involves simply taking the output of the committee to be the
average of the outputs of the individual networks. The members of the
committee are competing solutions of the classification problem, which
occurred as a result of starting the search in the parameter space
from different initial weights. It is also beneficial to combine
networks with different numbers of neurons in the hidden layer.

Finally, each new lightcurve has to be pre-processed and the features
extracted have to be fed to the trained network, which is defined by
the most probable parameter vector of weights $w_{\rm MP}$. The output
of the network is $P(C_1|x, w)$, the probability that the lightcurve
belongs to the class $C_1$ or microlensing given the inputs $x$ and
the weights $w$. The output can therefore then be used to make a
decision as to which class the current datum belongs.  Usually, the
lightcurve is assigned to the class for which the posterior
probability is largest. For a two-class problem with equal priors this
implies a formal decision boundary at $y=0.5$.  Although usually
different classes do have roughly equal prior probabilities in the
training set, in reality this need not be the case. We can correct for
this by adjusting the outputs of the trained network using the ratios
of prior probabilities for each class. As we show in Appendix A, this
can be exploited to calculate the microlensing rate directly from the
neural network outputs. We can also allow for this by moving the
decision boundary and classifying objects as microlensing only if the
probability exceeds some higher threshold than the formal decision
boundary.

Once we have transformed the new input pattern into the posterior
probability, it is important to have an estimate of the error in the
output. The error arises through variance and through undersampling in
the parameter space during training.  The variance part of the output
error is easiest to deal with. It can be approximated by taking the
standard deviation of the output of a committee of neural
networks. The second part of the output error is more awkward, but can
can be approximated by a method originated by MacKay (1992b), which we
now explain.

There will always be regions in input space with low training data
density. Typically the network with parameters $w_{\rm MP}$ will give
over-confident predictions in such regions. A representative output
then will be an output averaged over the distribution of network
weights, namely
\begin{equation}
P(C_1|x, D) = \int P(C_1|x,w)p(w|D)dw.
\end{equation}
Here, $C_1$ is the class (in our case, microlensing), $x$ denotes the
inputs and $D$ the data in the training set.  This integration cannot
be performed analytically, but there is a simple approximation, namely
\begin{equation}
P(C_1|x,D) \approx f(k(s) a_{\rm MP}), \qquad k(s) =
\left(1+\frac{\pi s^2}{8}\right)^{-1/2}.
\end{equation}
Here, $f$ is the activation function, $s$ is the network variance and
$a_{\rm MP}$ is the activation of the output neuron given the most
probable distribution of weights (the one that is found during network
training). The network variance is calculated using the methods of
Section 10.3 of Bishop (1995). It can be shown that this marginalized
or moderated prediction always has a value closer to 0.5 (the formal
decision boundary in two-class problems) than the most probable
one. Marginalization always drives the output closer to the formal
decision boundary.

When any network is applied to real data after training, it is
confronted with more complex light curves which inevitably extend
beyond the data domain encountered during training.  We caution that
neural networks sometimes classify these in an unpredictable manner,
as this amounts to an extrapolation of the decision boundaries. Our
use of marginalized or moderated output guards against this, as
unexpected or unpredicted patterns are then driven back to the formal
decision boundary.

\section{A CASCADE OF NETWORKS}

Neural networks can be arranged sequentially in a cascade to perform
complicated pattern recognition tasks. Here, the lightcurve data are
examined first with neural networks which eliminate the contaminating
variable stars. Then, lightcurves successfully passing this first
stage are analysed anew with neural networks which eliminate
contaminating supernovae. Excellent microlensing candidates must pass
both stages.

\subsection{A Network to remove the Variable Stars}

\begin{figure}
\epsfxsize=9cm \centerline{\epsfbox{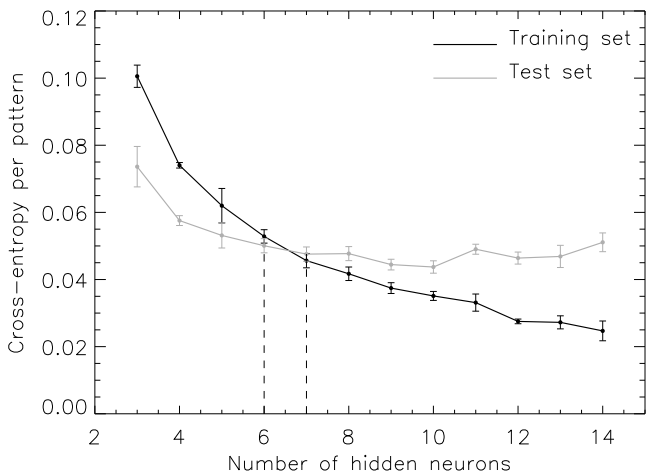}}
\caption{The standard cross-entropy error plotted against the number
of neurons in the hidden layer for the training set and the test set.
This begins to flatten for the test set data around 6 or 7 neurons.}
\label{fig:hid}
\end{figure}
\begin{figure}
\epsfxsize=9cm \centerline{\epsfbox{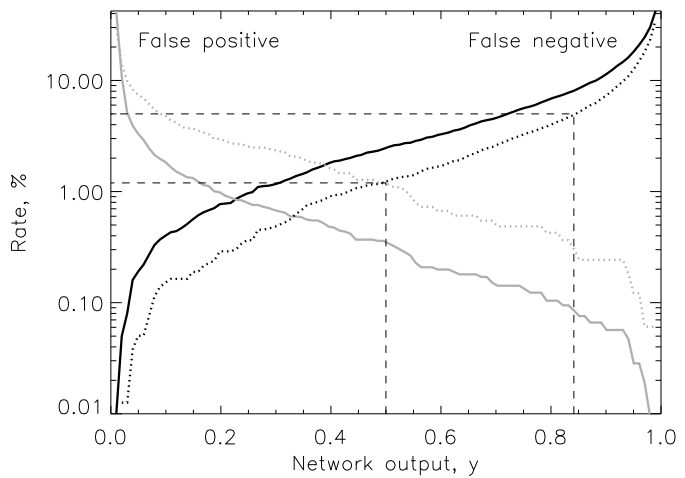}}
\caption{The false positive and false negative rates for single
passband data when the committee of neural networks is applied to the
test set. The horizontal axis is the network output. For the false
negatives, the vertical axis is number of misclassified microlensing
lightcurves {\it expressed as a percentage of the total number of
microlensing lightcurves.}  The solid line applies to the raw data
without any cleaning.  The dotted line corresponds to processing only
lightcurves with at least 5 datapoint with signal to noise greater
than 5 during the Einstein diameter crossing time.  For the false
negatives, the vertical axis is number of non-microlensing lightcurves
misclassified as microlensing {\it expressed as a percentage of the
total number of non-microlensing lightcurves.}  The solid line applies
to the raw data, while the dotted line corresponds to taking the
maximum of the output for the raw and the cleaned lightcurves.}
\label{fig:false}
\end{figure}
\begin{figure}
\epsfxsize=9cm \centerline{\epsfbox{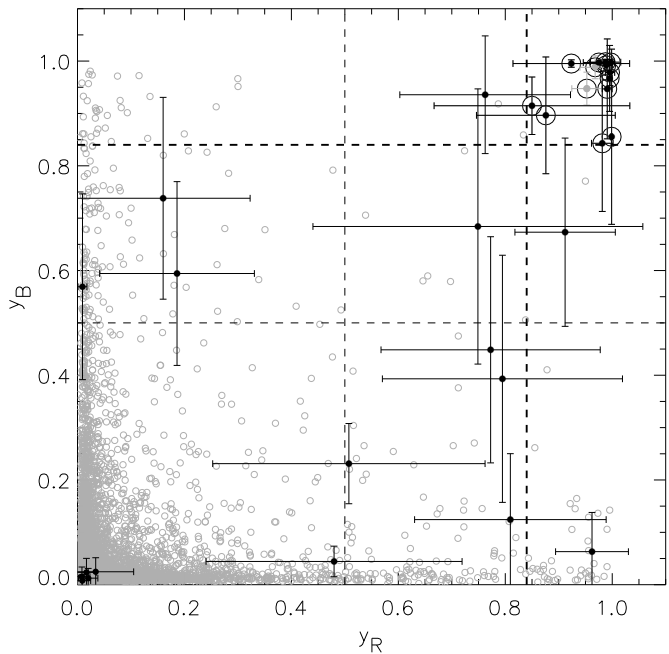}}
\caption{The locations of $\approx 22000$ MACHO lightcurves as given
by the outputs of the committee $y_R$ and $y_B$ on processing the red
data and the blue data respectively. These include the 29 lightcurves
that passed the loose selection of Alcock et al (2000), together with
$\sim 1000$ lightcurves in the vicinity of each candidate. Each point
gives the maximum of the moderated output for the raw and the cleaned
data, with the error bar giving the network scatter. A large open
circles around a point indicates that it lies above the decision
boundary ($y_R > 0.87$ and $y_B > 0.87$). Filled black dots represent
the 29 lightcurves selected by Alcock et al. (2000), while all other
lightcurves are represented by open grey dots.}
\label{fig:manytiles}
\end{figure}
\begin{figure}
\epsfxsize=\hsize \centerline{\epsfbox{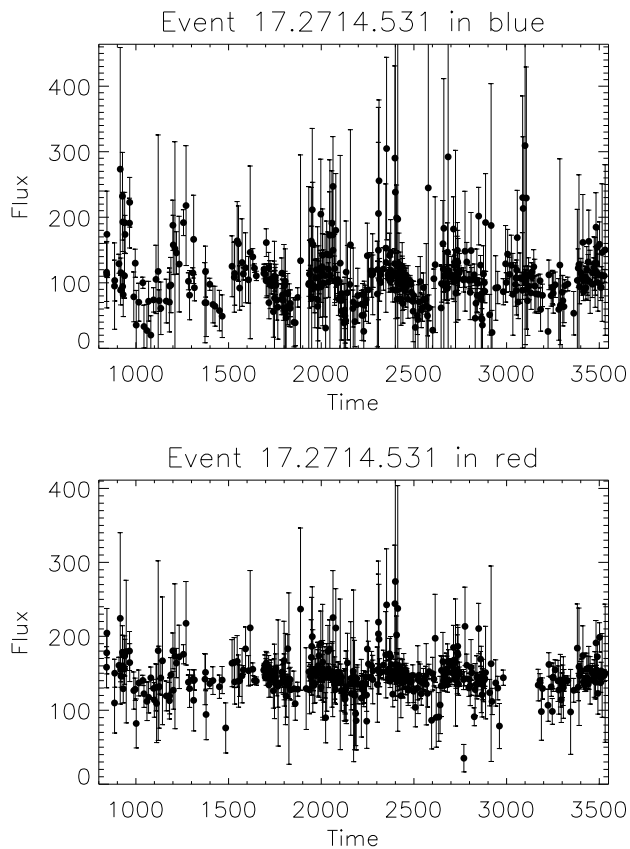}}
\caption{The lightcurve of one of the false positives. This
is close to the noise/microlensing border in parameter space.}
\label{fig:fpos}
\end{figure}

To eliminate the variable stars, we use the techniques developed in
Belokurov et al. (2003), but we make some modifications to the
training procedures. The training set contains 3513 patterns, 1500 of
which are derived from simulated microlensing lightcurves. These
events are generated by randomly choosing an impact parameter, an
Einstein crossing time between 7 days and 365 days and a time when the
event reaches maximum.  Random Gaussian noise is added to all the
lightcurves and the experimental sampling is used.  Only those events
that have 3 or more datapoints during the event with a signal-to-noise
greater than 5 are included in the training set.  The remaining 2014
lightcurves in the training set are broken down according to
Table~\ref{table:tset}. The sources of many of the variable star
lightcurves, such as Miras, novae and eclipsing variables, are derived
from the long data sequences provided by the American Association of
Variable Star Observers (AAVSO). Long period Cepheids are constructed
from their Fourier coefficients (e.g., Antonello \& Morelli 1996).
Artificial bumper lightcurves of a simple sinusoid shape with period
chosen randomly around the experiment lifetime are also used. The
period of a bumper is so long that typically only one bump is in the
dataset. In addition, 531 lightcurves randomly selected from the
MACHO database are included in the training set.

All the lightcurves are subjected to a spectral analysis to extract
parameters which are the inputs to the neural networks. Belokurov et
al. (2003) already devised 5 parameters, based on the underlying
premise that microlensing events are single, symmetric, positive
excursions from the lightcurve baseline. The same parameters are used
here.

All networks are trained using the Netlab package (Nabney 2002). The
optimization method is the variable metric or quasi-Newton algorithm
with Broyden-Fletcher-Goldfarb-Shanno updates (see Press et al. 1992;
Nabney 2002). The optimization is performed several times in sequence
with values of fractional tolerance decreased from $10^{-3}$ to
$10^{-6}$ by repeatedly halving. At the end of each convergence loop,
the hyper-parameter $\alpha$ is adjusted (according to eq. (2.4) of
MacKay (1992a) or eq. (10.74) of Bishop (1995)).

To find the optimal network architecture, we compare different
solutions with between 3 and 14 hidden neurons on both the training
set and the test set.  The latter set comprises 10000 simulated
microlensing lightcurves with noise and MACHO experimental sampling
and 10000 non-microlensing events (variable stars and lightcurves
drawn from MACHO LMC field 82 which has no candidate events).  The
cross-entropy error (see Bishop 1995, chap. 6) divided by the number
of patterns for the training and test sets is shown in
Figure~\ref{fig:hid} as a function of the number of hidden units. The
cross-entropy error per pattern for the training set slowly declines
with increasing number of neurons, but it begins to flatten
at about 6 or 7 hidden neurons for the test set.  Thus, we
choose to combine networks with 6 and 7 hidden units to form a
committee comprising in total 50 networks.

The committee is then applied to the test set to estimate the rate of
false negatives (microlensing events misclassified as not
microlensing) and false positives (non-microlensing events
misclassified as microlensing). Note that the probabilities or rates
of false negatives (or positives) are normalised to the total number
of microlensing (or non-microlensing) lightcurves respectively.  The
results for the raw data are shown in Figure~\ref{fig:false} in
unbroken lines.  The rate of false positives and false negatives are
equal with a value of $0.8 \%$ at a decision boundary of $y \approx
0.2$. However, most of the false negatives (genuine microlensing
lightcurves with an output $y <0.2$) have less than 5 datapoints
during the event with a signal-to-noise ratio $> 5$. If we process
only microlensing events with 5 or more such datapoints during the
events, then the false negative rate is shown as the black dotted line
in Figure~\ref{fig:false}.  In fact, the MACHO collaboration applies a
series of cuts to the raw data before analysis, which removes outliers
prevalent in the data. To mimic this, we ``clean'' the raw data using
the methods described by Belokurov et al (2003). If we process both
raw and clean lightcurves, taking the maximum output of the two, then
the false positive rate is increased as shown by the grey dotted
line. A decision boundary corresponding to the point where the two
dotted lines cross is $y = 0.5$. The false positive and false negative
rates in the test set are then both equal to $\sim 1\%$ for single
passband data.

In practice, we can choose to be more or less conservative. In other
words, we can reduce the incidence of false positive detections at the
expense of increasing the rate of false negative detections, or vice
versa. Where we choose this balance is controlled by the positioning
of the decision boundary.  As the MACHO data are taken in both blue
and red passbands, the network is actually applied twice.  For
classification as microlensing, an event must pass in both
passbands. Suppose the decision boundary corresponds to the false
negative rate $P$ for single passband data. This means that --
assuming that the distributions for each network are independent --
the false positive rate for data in two passbands is $\sim P^2$ and
the false negative rate is $\sim 2P$.  We select $P$ by insisting that
the number of false negatives in the entire MACHO dataset is $\lta 1$,
Using the information that -- as judged from the theoretical optical
depth -- the expected number of microlensing events in the entire
MACHO dataset is $O(10)$, this yields $P = 0.05$ which from the dotted
curve in Figure~\ref{fig:false} gives a decision boundary at
$y=0.84$. This we adopt in the rest of the paper.  It corresponds to a
false positive rate of $0.3 \%$

This choice of decision boundary gives rise to a negligible number of
bona fide microlensing events that are classified as
non-microlensing. Note that because non-microlensing is overwhelmingly
more common than microlensing, there will be more false positives than
false negatives.

To illustrate this, Figure~\ref{fig:manytiles} shows the locations of
$\approx 22000$ MACHO lightcurves.  The data for the red and blue
passbands are processed separately to give outputs $y_R$ and $y_B$.
Again, the value of the output that is plotted is the maximum of the
two outputs for the raw and the cleaned lightcurves.  The error bars
give the standard deviation of all the committee outputs.  The
decision boundary is shown in the bold broken line -- convincing
microlensing candidates have $y_{R,B} > 0.84$.  The 29 candidate
lightcurves identified by Alcock et al. (2000) are denoted by filled
black dots, while all other lightcurves are shown as open grey dots.
The outputs for Alcock et al.'s 29 candidates are recorded in the
first two columns of Table~\ref{table:Macho} and discussed in detail
in Section 4. Twelve of these 29 lightcurves satisfy $y_{R,B} > 0.84$,
namely 1a, 1b, 5, 6, 10a, 11, 14, 21-25. There are additionally 2
false positives (with MACHO lightcurve numbers 17.2221.1377 and
17.2714.531) with $y_{R,B} > 0.84$.  The lightcurves of one of the
false positives is illustrated in Figure~\ref{fig:fpos}.  Both have a
very low value for $x_1$ (the first input) and so they lie close to
the noise/microlensing border in parameter space.

Figure~\ref{fig:manytiles} can be used to illustrate the effects of
moving the decision boundary and therefore to assess the robustness of
our results. Suppose the decision boundary were to be relocated to
$y_R > 0.5$ and $y_B > 0.5$.  We expect this to reduce the numbers of
false negatives, at a cost of increasing the false positives. We now
find that there are 9 false positives, 7 of which lie close to the
noise/microlensing border. Additionally, there is one false positive
that lies in an undersampled region of parameter space, and one that
corresponds to a likely bumper. The gain is that a further 3
lightcurves are classified as microlensing (although these represent
only 2 additional events).

\subsection{A Network to remove the Supernovae}
\begin{figure*}
\epsfxsize=17cm \centerline{\epsfbox{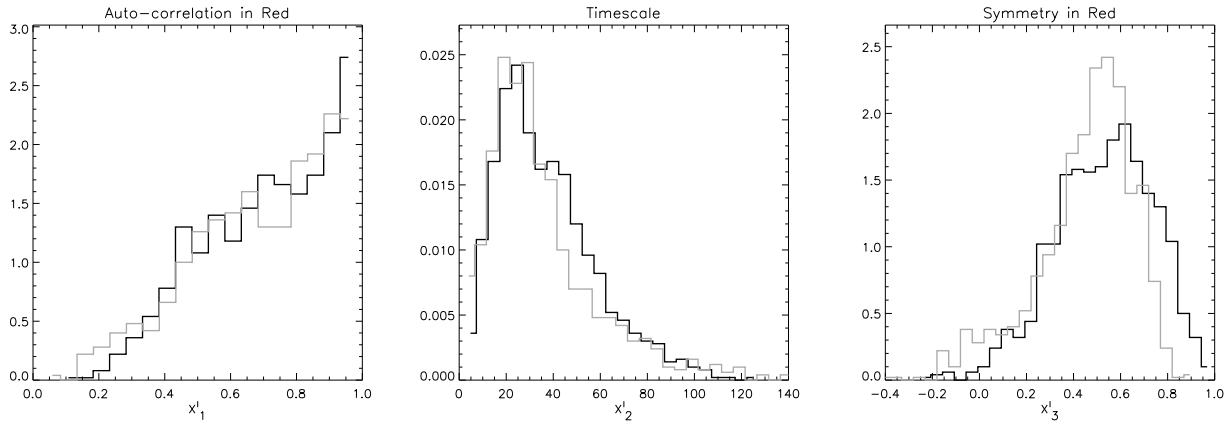}}
\caption{This shows the distributions of lightcurve shape features for
microlensing (black) and supernovae (grey) in the training set. The
timescale $x'_2$ is shown in days, while the auto-correlation
coefficient $x'_1$ and the symmetry measure $x'_3$ are dimensionless.}
\label{fig:shapefeatures}
\end{figure*}

To distinguish microlensing from supernovae occuring in background
galaxies is more problematic, as clearly pointed out in Alcock et
al. (2000). This is the job of the next network in the cascade. 

Gravitational microlensing of a point-source on a point-mass dark lens
moving with a constant velocity produces a symmetric brightness change
due to distortion of spacetime near the mass. A supernova lightcurve
is generated by an exploding star and is characterised by a very quick
rise followed by a steady decline. Based upon this knowledge, we might
hope to use the symmetry of the lightcurve as a discriminant
feature. However, microlensing lightcurves can appear much less
symmetric when the observational campaign has irregular time sampling
or when the beginning or end of the event is missed. On the other
hand, supernova lightcurves can seem symmetric if only the top part
of the lightcurve is sparsely sampled. This happens because distant
supernovae are generally faint objects and only briefly enter the
magnitude range of the survey.

Colour evolution during the event is another important discriminant.
The colours change dramatically during a supernova explosion as a
result of complicated radiation processes inside the ejecta. After a
fairly constant pre-maximum epoch with $B-V \approx 0$, a supernova of
type Ia typically starts turning red at the time of the maximum light,
it reaches $B-V \approx 1$ in about 30 days and then drops back (see
e.g., Phillips et al. 1999). This can be contrasted with the colour
behaviour during gravitational microlensing. Gravity bends light
irrespective of its frequency. Therefore, colour does not change
during microlensing. However, the achromaticity of the lightcurve only
holds good if the source star is resolved and the lens is dark. The
presence of other stars within the centroid of light or lensing by a
luminous object will result in a colour change during the event. At
the baseline, the colour is defined by the combined flux from all the
sources. The amplified star will contribute most of the colour around
the peak. The colour of a microlensing event can become redder or
bluer, depending on the population of the blend, but it usually
changes symmetrically about the peak with substantial correlation
between passbands (see e.g., Di Stefano \& Esin 1995, Buchalter,
Kamionkowski, \& Rich 1996).

Again, we build a training set with patterns of features extracted
from simulated microlensing and supernova lightcurves. Then, a
committee of networks is trained and applied to the lightcurves of all
transients found at the first stage of the data-mining.  In the
training set, simulated microlensing lightcurves have a slightly
different timescale distribution as compared to Section 3.1. The value
of the Einstein diameter crossing time is drawn from a Gaussian
distribution with zero mean and standard deviation of 75 days. This is
done to ensure that the set is dominated by fast transients, for which
confusion with supernova lightcurves is most problematic. Blending is
also added by changing the amplification to $(1- f_{B, R}) + A f_{B,
R}$ , where $A$ is the unblended amplification and the blending
fractions in blue and red passbands $f_{B, R}$ are drawn from a
Gaussian distribution with unit mean and standard deviation of 0.4.

We generate supernova lightcurves of type Ia only, as they are the
most luminous and hence should be the dominant contaminant in any
sample. For the templates, we use $R$ and $B$ passband data of
supernova SN 1991T from Lira et al. (1998). This is an unusually
bright supernova; however our algorithm chooses a random magnitude at
maximum so only the shape of the lightcurve is important. The $R$ and
$B$ colours from Lira et al. do not match MACHO passbands exactly
since MACHO imaging was performed in non-standard red ($\lambda
\lambda$ 5900-7800 \AA) and blue ($\lambda \lambda$ 4370-5900 \AA)
filters. This should not be a serious concern since the training set
data-cloud is smoothed by noise and irregular sampling. The simulated
supernova lightcurve is a randomly chosen part of the top of the
supernova template. We allow for extinction in the host galaxy by
permitting the lightcurves in the blue and red passbands to have
slightly different amplitudes.  The total detected brightness change
in magnitudes is $2.5 \log [(u^2+2)/u \sqrt{u^2+4}]$, where $u$ is
distributed uniformly between 0 and 1. In this way, the typical signal
in the subset of supernovae events in the training set correlates with
the typical signal in subset of microlensing events. All the
lightcurves have Gaussian noise added and are sampled with actual
MACHO sampling.
\begin{figure*}
\epsfxsize=17cm \centerline{\epsfbox{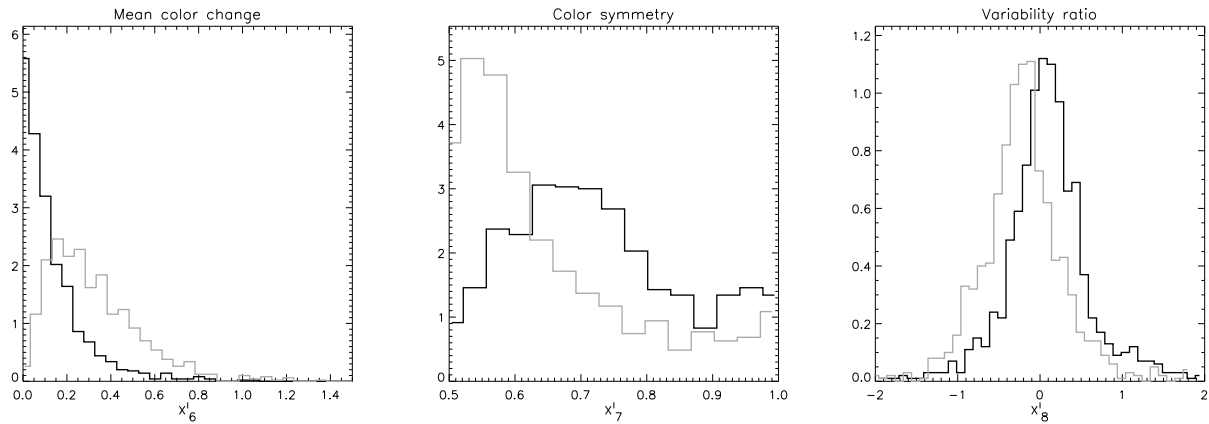}}
\caption{This shows distributions of colour features for microlensing
(black) and supernovae (grey) in the training set. Mean colour change
$x'_6$ is in magnitudes, $x'_7$ is mapped from (0,$\infty$) to (0.5,1)
with the sigmoid function and $x'_8$ is in logarithmic measure.}
\label{fig:colorfeatures}
\end{figure*}
To describe the shape of the lightcurve, we extract the following
features. First, $x'_1$ is the maximum value of the autocorrelation
coefficient. It can be regarded as a measure of the signal in the
lightcurve. To make the feature extraction more robust, we take
advantage of the fact that the lightcurve has already passed the first
stage of classification. So we can assume that the epoch of the
maximum light has been estimated by the first neural network. Thus,
the second feature $x'_2$ is the time between the peak and the instant
when the amplification exceeds $1.34$. For microlensing events, an
amplification of $1.34$ or greater means that projected position of
the source lies within an Einstein radius of the lens, and so $x'_2$
is exactly half the event duration. For supernova lightcurves, this
feature is well-defined, but does not correspond to anything with a
simple physical meaning.  The third feature $x'_3$ is the value of the
cross-correlation of the lightcurve with the time-reversed lightcurve
evaluated at lag $T$. Here, we use only the data-points within a
timescale $x'_2$ of the maximum in both the forward and backward
directions (the Einstein diameter crossing time for microlensing).
The lag $T$ is defined as the time difference between the instants
of maximum brightness of the lightcurve and the time-reversed
lightcurve.  The parameters $(x'_1, x'_2, x'_3)$ are all extracted
from the red lightcurve.  The fourth $x'_4$ and the fifth $x'_5$
features are the autocorrelation and symmetry parameters extracted
from the blue lightcurves.

Additionally, we feed the network with features characterizing the
colour change during the event. Note, that when the signal to noise of
the transient is low or when the colour change is minuscule, then the
error propagation might result in the destruction of any colour
signal. In other words, any signal in the colour is noisier than the
corresponding signal in the red or blue passbands separately.
Irregularity of the time sampling can further aggravate the problem,
since not all the measurements are taken simultaneously in both
colours. To account for this and to stabilize the colour, we extract
all the following features from lightcurves binned with a time
bin-size of 2 days. To estimate the total colour change during the
event, we calculate the weighted average excursion from the colour
baseline:
\begin{equation}
x'_6 = \left. \sum_{i=1}^n \frac{|(B-R)_i-(B-R)_0|}
{\sigma^2_{B-R, i}} \right/ \sum_{i=1}^{n} \frac{1}{\sigma^2_{B-R, i}}.
\end{equation}
Here, the index $i$ runs through all measurements within the Einstein
diameter crossing time and the baseline $(B-R)_0$ is the weighted
average colour outside the Einstein crossing time. The next feature
$x'_7$ is the ratio of total weighted absolute colour change before
and after the maximum light. This tests the symmetry of the colour
signal. For microlensing, this ratio takes values around 1, while for
supernova lightcurves it is close to zero. Therefore, we magnify the
range between 0 and 1 by transforming the ratio with the sigmoid
function. Finally, the last colour feature is the variability ratio as
defined by Welch \& Stetston (1993). It is the ratio of the total
normalized magnitude residuals in the blue and red filters, namely
\begin{equation}
x'_8 = \left. \sum_{i=1}^n |\delta B| \right/ \sum_{i=1}^n |\delta R|,
\end{equation}
where
\begin{equation}
\delta B = \frac {B_i - \overline B}
{\sigma_{B_i}}, 
\qquad
\delta R = \frac {R_i - \overline R}
{\sigma_{R_i}}, 
\end{equation}
Here the weighted means $\overline B, \overline R$ are calculated over
all epochs outside the Einstein crossing time. We take the logarithm
of the variability ratio so as to compress the range. Supernovae
lightcurves have, on average, smaller values of $x'_8$ than
microlensing lightcurves.

The distributions of lightcurve shape features are shown in
Figure~\ref{fig:shapefeatures}. It is clear from the first two panels
that $x'_1$ and $x'_2$ serve as control features. The autocorrelation
and timescale distributions of supernova and microlensing lightcurves
do not differ much. This is reassuring since it indicates that we are
probing similar signal regimes of the two different variability
classes.  The distribution of $x'_3$ (the third panel of
Figure~\ref{fig:shapefeatures}) confirms our choice of this feature as
a symmetry measure with microlensing dominating around values of
$\gta 0.8$.

Figure~\ref{fig:colorfeatures} shows the distribution of colour
related parameters. From the first panel, it follows that, as
expected, the amplitude of the colour change is significantly lower
for microlensing. Note, however, that there is a tail in the $x'_6$
distribution that stretches as far as 1.5 magnitudes for both
microlensing and supernovae. The colour signal looks very symmetric
for microlensing with $x'_7$ peaking at $\sim 0.7$. Let us recall that
the original colour symmetry ratio was transformed with the sigmoid
function, which means that 1 is mapped onto value $\simeq 0.73$. The
distribution of $x'_7$ for the supernova lightcurves peaks around
0.55, which corresponds to a value of 0.2 in the symmetry
ratio. Finally, the variability ratio $x'_8$ is presented in the third
panel of this figure. The mean value of the logarithm of $x'_8$ for
microlensing is zero and the distribution itself is symmetric, while
supernovae prefer smaller values of this feature, typically by factor
of $10^{-0.2} \simeq 1.6$.

The total number of patterns in the training set is 2000, one half are
extracted from microlensing lightcurves and the other half from the
simulated lightcurves of supernovae. For networks with more than 5
neurons, the data misfit keeps decreasing monotonically. We therefore
choose to use 10 networks with 5 hidden units to form the committee.
The candidate microlensing events towards the LMC are then processed
with the network and the outputs recorded in the third column of
Table~\ref{table:Macho}. The output $y'$ can be interpreted as the
probability that the lightcurve is not a supernova. The optimum
decision boundary can be found by examining the false positive and
negative rates as in Section 3.1; however, for the purposes of this
paper, it suffices to interpret $y'\ll 0.5$ as a strong supernova
candidate, $y' \gg 0.5$ as definitely not a supernova, and $y' \approx
0.5$ as indeterminate.

\begin{table}
\begin{center}
\begin{tabular}{l|c|c|c}\hline \hline
Event & $y_R$ & $y_B$ & $y'$ \\ \hline
1a & 0.88 $\pm$ 0.13 & 0.90 $\pm$ 0.11 & 0.97 $\pm$ 0.01 \\
1b & 0.99 $\pm$ 0.01 & 0.98 $\pm$ 0.03 & 0.95 $\pm$ 0.01 \\
4 & 0.81 $\pm$ 0.18 & 0.12 $\pm$ 0.13 & 0.90 $\pm$ 0.02 \\
5 & 0.99 $\pm$ 0.002 & 0.86 $\pm$ 0.17 & 0.74 $\pm$ 0.18 \\
6 & 0.98 $\pm$ 0.03 & 0.99 $\pm$ 0.003 & 0.97 $\pm$ 0.02\\
7a & 0.77 $\pm$ 0.21 & 0.45 $\pm$ 0.22 & 0.84 $\pm$ 0.10\\
7b${^*}$ & 0.02 $\pm$ 0.02 & 0.02 $\pm$ 0.02 & 0.21 $\pm$ 0.05\\
8 & 0.51 $\pm$ 0.25 & 0.23 $\pm$ 0.08 & 0.86 $\pm$ 0.04\\
9$^{*, {\rm binary}}$ & 0.76 $\pm$ 0.16 & 0.94 $\pm$ 0.11 & 0.67 $\pm$ 0.13\\
10a$^{{\rm SN}}$ & 0.85 $\pm$ 0.18 & 0.92 $\pm$ 0.05 & 0.82 $\pm$ 0.12\\
10b$^{{\rm SN}}$ & 0.16 $\pm$ 0.16 & 0.74 $\pm$ 0.19 & 0.88 $\pm$ 0.01\\
11$^{*, {\rm SN}}$ & 0.98 $\pm$ 0.02 & 0.84 $\pm$ 0.13 & 0.05 $\pm$ 0.01\\
12a$^{{\rm SN}}$ & 0.96 $\pm$ 0.07 & 0.06 $\pm$ 0.07 & 0.01 $\pm$ 0.01\\
12b$^{{\rm SN}}$ & 0.75 $\pm$ 0.31 & 0.68 $\pm$ 0.26 & 0.42 $\pm$ 0.25\\
13 & 0.03 $\pm$ 0.07 & 0.03 $\pm$ 0.03 & 0.96 $\pm$ 0.04\\
14 & 0.92 $\pm$ 0.11 & 0.99 $\pm$ 0.007 & 1.00 $\pm$ 0.00\\
15 & 0.01 $\pm$ 0.01 & 0.01 $\pm$ 0.01 & 0.84 $\pm$ 0.03\\
16$^{*, {\rm SN}}$ & 0.01 $\pm$ 0.01 & 0.57 $\pm$ 0.18 & - \\
17$^{*, {\rm SN}}$ & 0.01 $\pm$ 0.01 & 0.02 $\pm$ 0.02 & 0.04 $\pm$ 0.01\\
18 & 0.91 $\pm$ 0.09 & 0.68 $\pm$ 0.18 & 0.95 $\pm$ 0.03\\
19$^{*, {\rm SN}}$ & 0.02 $\pm$ 0.02 & 0.01 $\pm$ 0.02 & 0.07 $\pm$ 0.06\\
20$^*$ & 0.8 $\pm$ 0.22 & 0.39 $\pm$ 0.24 & 0.23 $\pm$ 0.20\\
21 & 0.99 $\pm$ 0.03 & 0.99 $\pm$ 0.02 & 1.00 $\pm$ 0.00\\
22 & 0.99 $\pm$ 0.001 & 0.99 $\pm$ 0.002 & 0.98 $\pm$ 0.01\\
23 & 0.99 $\pm$ 0.01 & 0.99 $\pm$ 0.01 & 0.96 $\pm$ 0.01\\
24$^{*, {\rm SN}}$ & 0.99 $\pm$ 0.005 & 0.97 $\pm$ 0.06 & 0.61 $\pm$ 0.26\\
25 & 0.99 $\pm$ 0.01 & 0.95 $\pm$ 0.1 & 0.98 $\pm$ 0.01\\
26$^{*, {\rm SN}}$ & 0.19 $\pm$ 0.14 & 0.59 $\pm$ 0.18 & 0.87 $\pm$ 0.02\\
27$^*$ & 0.48 $\pm$ 0.24 & 0.04 $\pm$ 0.03 & 0.70 $\pm$ 0.01\\ 
\hline
\end{tabular}
\end{center}
\caption{This shows the output of the committee of neural networks
(the posterior probability of microlensing) on the set of candidates
towards the LMC identified by Alcock et al. (2000).  Stars mark events
that did not pass MACHO's selection criteria A. A superscript SN marks
a supernova as judged by Alcock et al. (2000). The first two columns
($y_R$ and $y_B$) are the outputs of the network of Section 3.1 on the
red and blue data, the third column ($y'$) is the output of the
network of Section 3.2.}
\label{table:Macho}
\end{table}

\section{New light on the MACHO candidates} 

First, let us recall that Alcock et al. (2000) used a series of
conventional cuts to identify microlensing events. The set A selection
criteria are ``designed to accept high quality microlensing
candidates''. The set B criteria are ``designed to accept any light
curves with a significant unique peak and a fairly flat baseline''.
19 lighcurves pass the set A criteria and 29 pass the looser set
B. Sometimes the same source star has two lightcurve because, for
example, it lies in an overlap region. Eight of the 29 lightcurves
(1a, 1b, 7a, 7b, 10a, 10b, 12a and 12b) correspond to just four stars.
Finally, Alcock et al. apply a supernova cut, insisting that a blended
microlensing lightcurve is a better fit than a SN Ia template.  This
finally leaves 13 events in set A (events 1, 4-8, 13-15, 18, 21, 23
and 25) and 17 events in set B (everything in set A together with 9,
20, 22 and 27). Subsequently, event 22 was confirmed to be a Seyfert
galaxy and so can be removed from set B (Sutherland, private
communication).

\subsection{Microlensing versus Variable Stars} 

\begin{figure*}
\epsfxsize = 17cm \centerline{\epsfbox{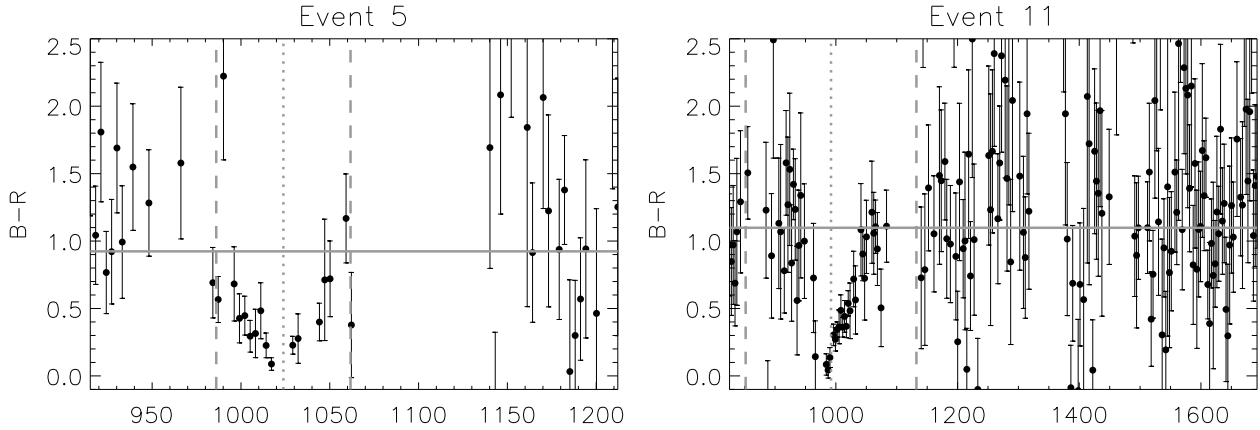}}
\caption{This shows the colour evolution of events 5 and 11. Event 5
shows a colour shift that changes symmetrically about the peak in the
flux of the event. This is characteristic of blended microlensing
events. Event 11 is a supernova candidate, as evidenced by the stable
colour in the pre-maximum epoch, the rapid reddening at maximum,
followed by the colour becoming increasingly blue. The classical
supernova colour curve as depicted in Lira et al. (1998) is shifted
because of extinction in the host galaxy.  The vertical axis is $B-R$
in magnitudes and the horizontal axis is time in JD-2448000. The
dotted vertical line is the peak of the event, while the dashed
vertical lines mark the time over which the amplification exceeds
$1.34$.}
\label{fig:lmc_color}
\end{figure*}
\begin{figure*}
\epsfxsize = 17cm \centerline{\epsfbox{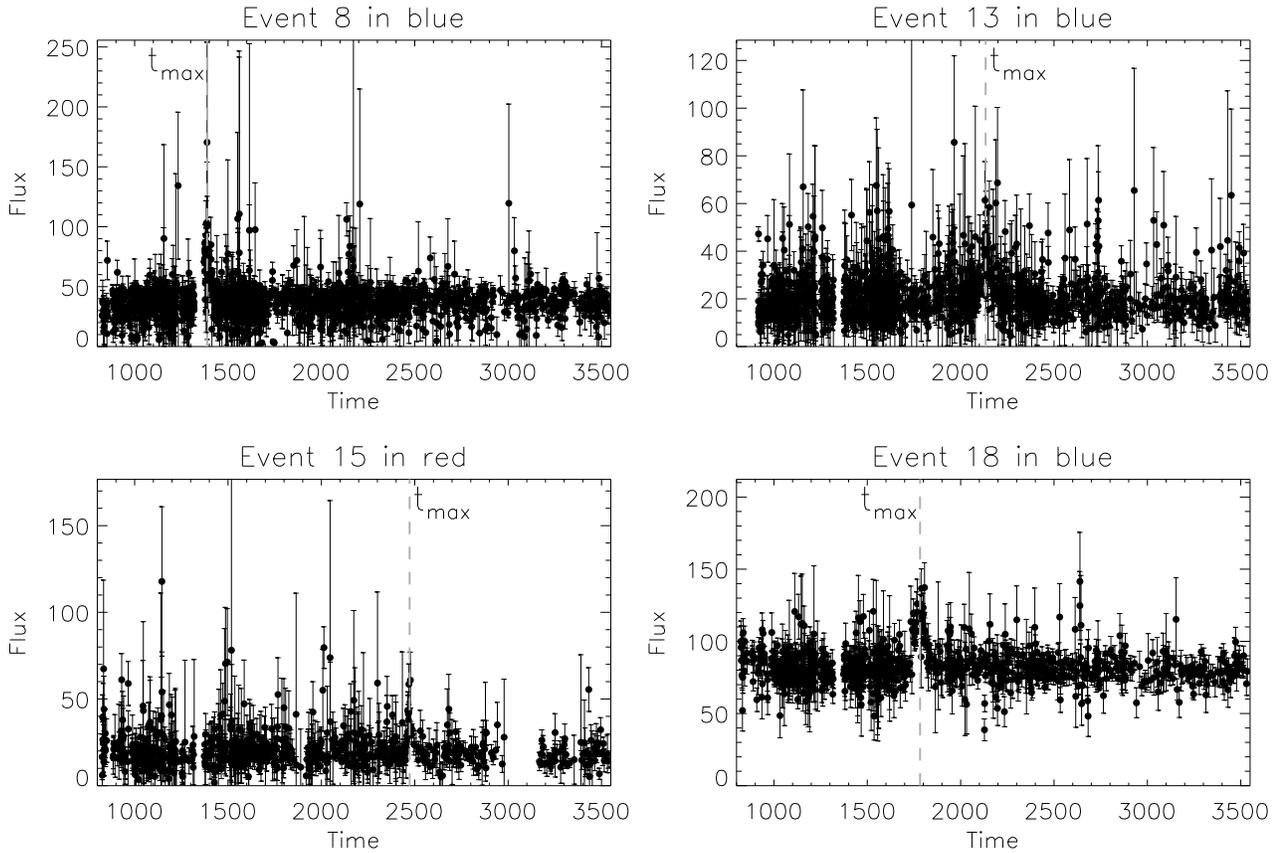}}
\caption{This shows the lightcurves for 4 events which received low
probability values $y$ in one or both filters. These are all included in
set A of Alcock et al. (2000) of convincing microlensing candidates,
but are not confirmed by our neural network analysis.  The vertical
axis is flux in ADU s$^{-1}$ and the horizontal axis is time in
JD-2448000. Vertical lines mark the peak of the event.}
\label{fig:lmcmiss}
\end{figure*}

Table~\ref{table:Macho} shows the predictions of committees of neural
networks for the LMC microlensing candidates selected by MACHO. First,
let us concentrate on the output in the first two columns which is
provided by the committee of neural networks to eliminate variable
stars (see section 3.1).  Let us recall that the output $y$ is the
posterior probability of microlensing.

In total, 7 out of 13 candidates from MACHO set A receive $y>0.84$ in
both red and blue filters: 1, 5, 6, 14, 21, 23, 25. These events can
be regarded as secure microlensing identifications.

Six events from MACHO set A fail the test for microlensing. Event 18
is a marginal case, as it is identified in the red ($y_R = 0.91$) but
not in the blue ($y_B = 0.68$). It is a low signal-to-noise event,
with one of the smallest maximum amplifications $A_{\rm
max}=1.54$. Events 4, 7, 8, 13 and 15 have $y < 0.84$ in both
bands. Some of these lightcurves are noisy with no stable baseline,
such as events 13 and 15. Event 8 has an apparently asymmetric shape,
partly because the beginning of the event is lacking due to a gap in
the observational campaign. The lightcurves of some of the failed
events are shown in Figure~\ref{fig:lmcmiss}.

One of the lightcurves that was selected by MACHO as a result of
applying only the loose selection criteria B gets an output
$y_{R,B}>0.84$.  This is event 22. The remaining three candidates --
events 9, 20 and 27 -- all fail our microlensing test of $y_{R,B} >
0.84$.

The four supernova suspects as judged by MACHO (events 16, 17, 19, 26)
fail the microlensing network committee.  The other three candidates
also suspected by MACHO of being supernova lightcurves, 10a, 11 and
24, are classified with probability $y_{R,B}>0.84$. They are, however,
discarded after being tested with the second neural network committee.

\subsection{Microlensing versus Supernovae} 

Convincing supernova candidates must have an output $y'\ll 0.5$ from
the second neural network committee. There are five events satisfying
this, namely the 4 MACHO supernova suspects (11, 12a,b, 17, 19) plus
candidate 20. The colour evolution of event 11 is illustrated in
Figure~\ref{fig:lmc_color}.  Although not identified by MACHO as a
supernova candidate, event 20 has a typical supernova colour
evolution. MACHO claims there are four more supernovae in the dataset,
namely events 10a,b, 16, 24 and 26. Unfortunately, candidate 16 has no
information in the red filter, but it is classified as
non-microlensing in the blue colour by the first network.  Event 24
has probability $y'=0.6$, the error is large and -- to be conservative
-- we conclude its origin is unknown.

Candidates 10a,b and 26 have outputs greater than 0.8. These two
events have timescales of $\sim 40$ days. If they are indeed
supernovae, it means that signal is present only for $\sim 20$ days
after the maximum light. The colour reaches a maximum after $\sim 30$
days, but even after 20 days B-V can be as much as $0.5-1.0$ mag (see
Figure 1 in Phillips et al. 1999). Neither event 10a,b nor 26 shows any
significant colour change. Hence, we do not confirm the supernova
classification of Alcock et al. (2000).

There is just one candidate that has a substantial colour signal
identified as blended microlensing by the neural networks. This is
event 5. The colours evolve symmetrically during the event, which
becomes $\sim 1$ mag bluer. It has output $y'= 0.74$ and is
illustrated in Figure~\ref{fig:lmc_color}.

\subsection{Numbers of Events}

In conclusion, then, the committees of neural networks reckons that
there are 7 convincing microlensing candidates. These are the events
1, 5, 6, 14, 21, 23 and 25. All these events have an output that
always lies above the decision boundary $y>0.84$.  They also judged to
be not supernovae ($y' \gg 0.5$). Of the remaining events, 10a and 18
are possible, but not convincing, microlensing candidates.

Compared to Alcock et al.'s (2000) set A, we have discarded events 4,
7, 8, 13, 15 and 18 (which is a marginal case). Four of the events
that we have excised from Alcock et al.'s sample A are shown in
Figure~\ref{fig:lmcmiss}. In each case, we show the data from the
passband which yields the lowest network probability.  None of the
events in set B (events 9, 20 and 27) are identified as microlensing
by the committee, while event 22 is known to be a Seyfert on other
grounds\footnote{Note that event 22 would otherwise have been
classified as microlensing by the neural networks. No method can
classify event 22 as a Seyfert galaxy on the basis of the MACHO
photometry alone without the follow-up observations.}

Alcock et al. reckoned there were 8 supernova suspects. We confirm 4
of these (events 11, 12, 17, 19) and we also found 1 new one (event
20). The remainder of Alcock et al's supernova candidates are not
thought to be either convincing supernova or microlensing candidates
by the committees.

\subsection{Optical Depth}

How does this affect the optical depth results? In qualitative terms,
the optical depth must be significantly lower than the value of
$1.2^{+0.4}_{-0.3} \times 10^{-7}$ of Alcock et al. (2000) and more in
accord with the results of the EROS collaboration (Laserre et
al. 2000). This is because the number of convincing microlensing
candidates has been reduced from 17 to 7 in our analysis. However, in
quantitative terms, the optical depth is not so easy to compute
without re-processing the entire MACHO dataset of $\sim 11.9$ million
lightcurves. There may be lightcurves that the neural networks
identify as microlensing, even though MACHO did not. This seems
unlikely, as no new candidates emerged from the $\approx 22000$ MACHO
lightcurves we have re-processed. However, it cannot yet be ruled out,
and so we do not provide an estimate for the optical depth from our
neural networks. Here, we merely note that the number of events has
been roughly halved, and we speculate that a concomitant reduction in
the optical depth might be expected.

\section{Conclusions} 

This paper has demonstrated the power of machine learning techniques,
such as neural networks, for the classification of events in massive
variability datasets. Using the specific example of the microlensing
surveys, committees of neural networks have been devised to
discriminate against common forms of stellar variability and against
supernovae. The output of the neural network is the posterior
probability of microlensing, given the prior distribution in the
training set. The error on the probability can be straightforwardly
calculated.

The networks have been used to process some of the data ($\approx
22000$ lightcurves) taken towards the Large Magellanic Cloud by the
MACHO collaboration (Alcock et al. 2000).  The latter authors provide
a set of 13 events whose identification as microlensing is believed to
be secure and a further 4 events whose identification is possible. The
neural networks confirm the microlensing nature of only 7 of these
possible 17 events. 

Without processing the entire dataset ($\sim 11.9$ million
lightcurves), we cannot be sure that there are no events missed by
Alcock et al. (2000) which would be classified as microlensing by the
networks.  It is reasonable to argue that this is unlikely, as the
$\approx 22000$ MACHO lightcurves we have re-processed provide no new
candidates. But, this remains a plausible speculation rather than an
empirically derived fact.  Hence, we can only speculate that, as the
number of events has been roughly halved, so the optical depth will be
similarly reduced.

For comparison, Alcock et al. (1997) estimate the optical depths of
the thin disk, thick disk and spheroid to be $2.2 \times 10^{-8}$,
whilst the optical depth of the stellar content of the LMC to be $3.2
\times 10^{-8}$ on average. In other words, from the known stellar
populations in the outer Galaxy and the LMC, the optical depth must be
at least $5.4 \times 10^{-8}$. This may well be enough to provide the
7 events whose microlensing nature we confirm.

There is supporting evidence for the belief that the known stellar
populations are providing the bulk of the lenses both from the exotic
events and from the lensing signal towards the Small Magellanic Cloud
(SMC).  First, the exotic events yield additional information which
can break some of the microlensing degeneracies and thus give indirect
evidence on the location of the lens. There are two exotic events
towards the LMC and two towards the SMC (Bennett et al. 1996;
Palanque-Delabrouille 1998; Kerins \& Evans 1999; Afonso et al. 2000;
Alcock et al. 2001a; Evans 2002). In all cases, the exotic events
favour an interpretation in which the lens lies in the Magellanic
Clouds.  Additionally, Alcock et al. (2001b) imaged one of the events
towards the LMC and identified the lens as a nearby low mass disk
star.

Second, as Afonso et al. (2003a) point out, the duration of the events
towards the SMC is very different from the duration towards the LMC.
The EROS collaboration constrain the optical depth towards the SMC to
be $< 10^{-7}$ at better than the 90 \% confidence level, based on an
admittedly small sample. Both these facts militate against the idea
that a single population of objects in the Milky Way halo is causing
the microlensing events. The mass function, internal kinematics and
proper motions of the SMC and LMC are different, so that differences
in the distributions of microlensing events are expected if the lenses
lie predominantly in the Magellanic Clouds. Based on roughly spherical
models of the dark halo, the optical depth towards the SMC is expected
to be greater than that towards the LMC if the halo provides most of
the lenses. Hence, the paucity of events towards the SMC is beginning
to be highly problematic for halo interpretations of the events.

\section*{Acknowledgments}
VB and YLD thank PPARC for financial support. We are grateful to the
anonymous referee for a number of helpful comments.  In this research,
we have used, and acknowledge with thanks, data from AAVSO
International Database based on observations submitted to the AAVSO by
variable star observers worldwide.  This paper also utilizes public
domain data obtained by the MACHO Project, jointly funded by the US
Department of Energy through the University of California, Lawrence
Livermore National Laboratory under contract No. W-7405-Eng-48, by the
National Science Foundation through the Center for Particle
Astrophysics of the University of California under cooperative
agreement AST-8809616, and by the Mount Stromlo and Siding Spring
Observatory, part of the Australian National University.  We
particularly wish to thank David MacKay for advice on neural networks.

{}

\appendix

\section{Neural Network Estimators of the Microlensing Rate}

It is interesting to develop methods of calculating the theoretical
microlensing quantities directly from the outputs of neural networks.

Let us define $E(x)$ to be the ratio of the density of microlensing
events in the training set to the true density, i.e.,
\begin{equation}
E(x) = {P(x|C_1) \over \hat P(x|C_1)}
\end{equation}
Here $\hat P(x|C_1)$ is the conditional probability of microlensing
(i.e., class 1) in the real world.

The output of the neural network is the posterior probability, and
relies on the prior probabilities of different classes of variability
in the training set. As follows from Table~\ref{table:tset}, the
prior probability of microlensing in the training set is at least
$10^6$ times larger than that in the real world. Indeed, the training
set contains a large number of microlensing lightcurves to ensure a
good variety of training examples. Therefore, the outputs of the
trained neural network need to be adjusted with respect to the
real-world priors. It has been shown (e.g., Saerens et al. 2002) that
a simple iterative procedure can help to tackle the problem. For 
microlensing, it follows from Bayes' theorem that:
\begin{equation}
\label{bayesa}
\frac{P(C_1|x)P(x)}{P(C_1)}= E(x) \frac{\hat P(C_1|x) \hat P(x)}{\hat P(C_1)}
\end{equation}
For variable stars, the same equation holds good without the
correction for input space sampling, namely
\begin{equation}
\label{bayesb}
\frac{P(C_2|x)P(x)}{P(C_2)}=\frac{\hat P(C_2|x) \hat P(x)}{\hat P(C_2)}
\end{equation}
This assumes that the sampling never causes the misclassification of a
variable star as a microlensing event.  In our notation, quantities
with a hat superscript refer to the real world, whereas unhatted
quantities refer to the training set. Let us now recall that the
activation $a$ of the output neuron can be interpreted as a logarithm
of the ratio of posterior probabilities:
\begin{equation}
a \equiv \log \frac{P(C_1|x)}{P(C_2|x)}
\end{equation}
This is simply the consequence of using the sigmoid function for
activation. Applying formulae~(\ref{bayesa}) and (\ref{bayesb}) to
each of the two classes and taking the ratio of probabilities, we
easily obtain:
\begin{equation}
\label{correction}
\hat a = a(x)-\log\frac{E(x) P(C_1)}{\hat
P(C_1)}+\log\frac{P(C_2)}{\hat P(C_2)}
\end{equation}
Typically, $P(C_1)/{\hat P} (C_1) \sim 10^7$. If the activation was
originally $\lta 7$, then this transformation maps it to below the
decision boundary. Only if the output is originally $> 0.999$ does the
event remain above the decision boundary.

Thus, having initialized $P(C_k)$ by the frequencies of the classes in
the training set, we perform the following iterative steps. Firstly,
the formula
\begin{equation}
{\hat P} (C_1) \approx \frac{1}{N}\sum_{i} y_i
\end{equation}
is used to estimate the true probability of microlensing.  Here, $i$
runs through all $N$ patterns in the data set.  Then, for each pattern
in the data set activation $a_i$ is adjusted using
formula~(\ref{correction}) and the output $y_i$ is re-calculated. The
process is repeated until convergence. At the beginning of the
iteration, $P(C_1) / {\hat P} (C_1)$ is $0(10^7)$ so that the sampling
factor $E(x)$ does not play an important role. However, after a few
iterations, it becomes important.  $E(x)$ is really a higher
dimensional analogue of the temporal efficiency $\epsilon$. It can be
calculated by generating events with uniform priors. In every cell of
input space, we calculate the ratio of accepted events to generated
events.

The output of this procedure is the true probability of microlensing
in the experiment monitoring $N_\star$ stars and lasting for a
duration $T$. From this, the microlensing rate is
\begin{equation}
\Gamma = {N_\star {\hat  P}(C_1) \over T}
\end{equation}
The advantage of this algorithm is that the rate can be computed
directly from the dataset, without the intervening steps of candidate
selection and efficiency estimation.

\end{document}